# 20 ps Non-Destructive Read and 1 ns Write Operations at <5 V in Ferroelectric $HfO_2/ZrO_2$ Non-Volatile Memories

A. Baigol[1,*], R. Hamming-Green[2,3], P. Uriarte Vicandi[1], J. Gao[1], T. Zellweger[1], A. Panda[1], A. Emboras[1], M. Csontos[4], M. Luisier[1], B. Noheda[2,3], and L. Bégon-Lours[1]

[1]Integrated Systems Laboratory, ETH Zürich, Zürich, Switzerland; *Email: abaigol@ethz.ch; [2]Zernike Institute for Advanced Materials, University of Groningen, Groningen, Netherlands; [3]CogniGron (Groningen Cognitive Systems and Materials Center), University of Groningen, Groningen, Netherlands; [4]Institute of Electromagnetic Fields, ETH Zürich, Zürich, Switzerland

***Abstract*—Achieving low-voltage, nanosecond multi-level programming and non-destructive read-out of ferroelectric non-volatile memories (NVM) is critical for analog in-memory computing architectures relying on ferroelectric capacitive devices (FeCap). We integrate $HfO_2/ZrO_2$ ferroelectric nanolayers concurrently in the BEOL of CMOS and on $SiO_2/Si$, achieving nanosecond multilevel switching with programming voltages below 5 V. Partial ferroelectric switching enhances FeCap endurance above $10^{11}$ cycles, leading to MemCapacitance (MC) states with non-destructive read-out and 10-year retention. However, experiments reveal the collapse of the MC window for read frequencies above 1 MHz. To overcome this speed limit, we introduce a novel, non-destructive readout methodology. Using electrical pulses with widths down to 20 ps, below the RC time constant of the FeCaps, we enable measurement of the polarization-dependent leakage current, providing ultrafast and non-destructive read operations at only 14 fJ.**

## Introduction

Ferroelectric (FE) hafnia is a key enabler for CMOS-compatible, emerging non-volatile memories [1], particularly for in-memory computing architectures aimed at accelerating AI workloads [2]. The latter relies on the non-destructive read of analog devices, for example, by a parallel voltage drop through crossbar arrays of resistive weights [3]. Recently, MemCapacitive (MC) devices were proposed for energy-efficient neuromorphic hardware [6, 7]. FE MC technologies [6] exhibit faster switching speed than other MC technologies based on charge trapping [7], [8], but two challenges remain: first, the need for high programming voltages and programming endurance. Here, we engineer sub-micrometer-square FeCap devices based on an asymmetric stack, exhibiting nanosecond programming speed below 5 volts. We show that partial switching of FE domains allows multilevel MC programming and boosts the memory endurance. The second challenge is the read frequency: we show that, as in monoclinic hafnia [9] and in FE HZO solid solution [10], the non-destructive read of the MC window cannot be performed above 1 MHz. Working towards high-speed and energy-efficient FE memory devices, we propose a novel non-destructive read scheme using 20 picosecond pulses.

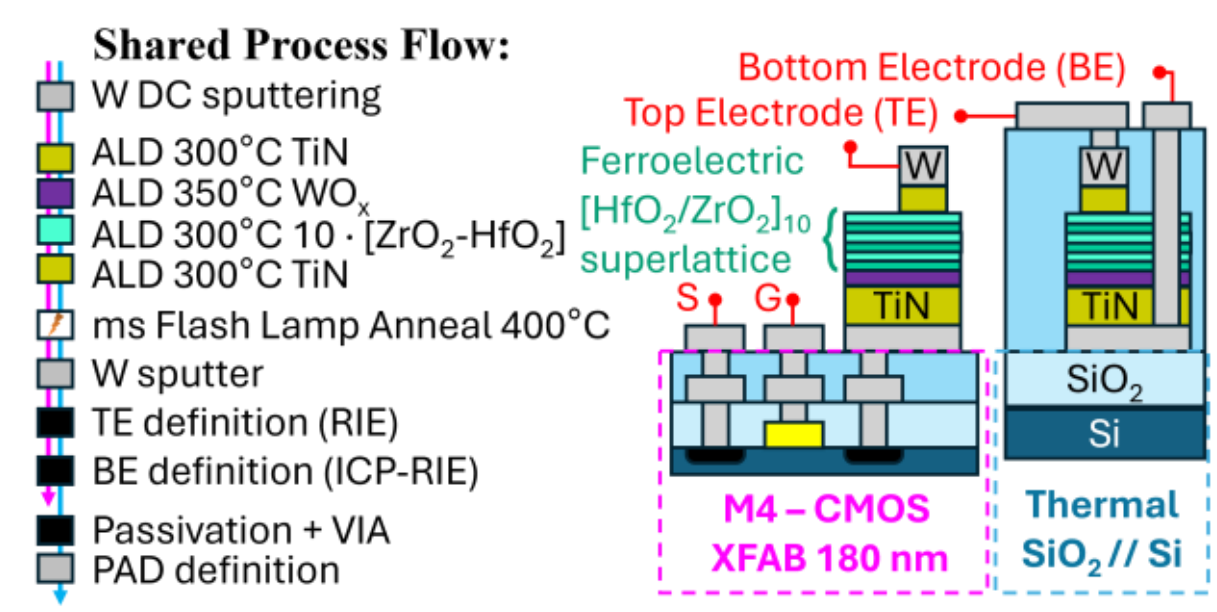

Fig. 1. Shared process-flow for the fabrication of FeCaps integrated in the BEOL of XFAB 180 nm CMOS and on $SiO_2$ with schematic cross-sections. Prior to ALD, CMOS wafers are treated with $O_2$ plasma.

## FeCap Fabrication

The functional stack consists of a 10 nm thick $[HfO_2/ZrO_2]_{10}$ nanolaminate (HZO) grown on $WO_x$ and TiN with Atomic Layer Deposition (ALD) (**Fig. 1**). It is co-deposited in the BEOL of XFAB 180 nm CMOS and on a thermal $SiO_2$ chip. Two aspects are novel compared to prior work [11]: first, we developed an e-beam lithography process to scale FeCaps down to 0.12 µm$^2$. Second, 50 nm of W were sputtered prior to the TiN electrode to improve the metal line conductivity.

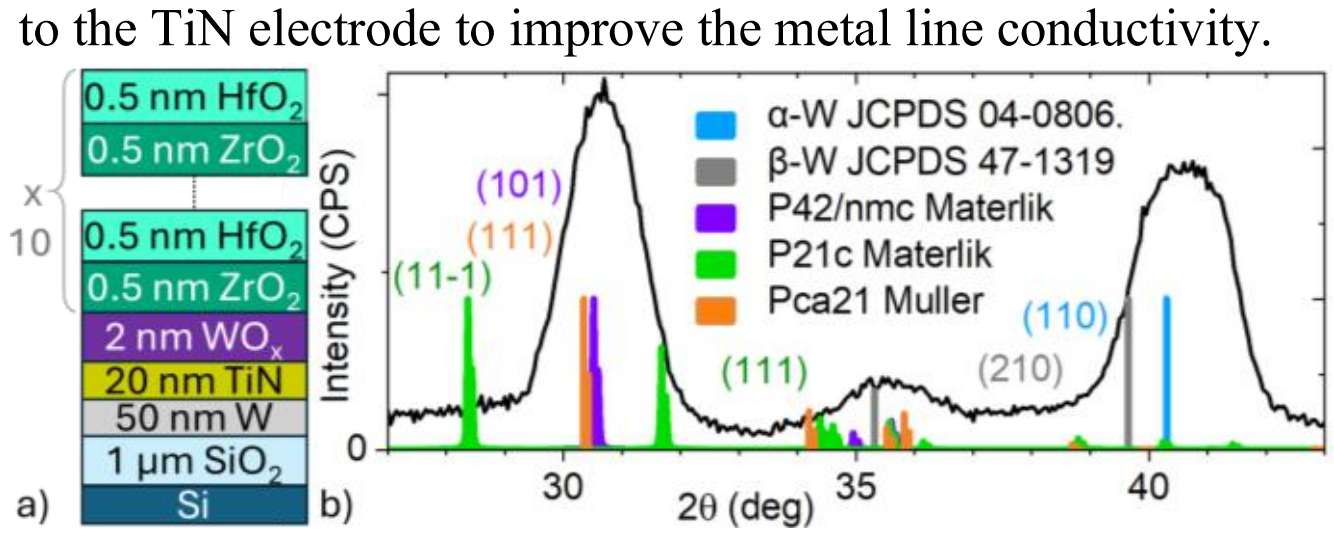

Fig. 2. a) $[HfO_2/ZrO_2]_{10}/WO_x/TiN/W/SiO_2//Si$ stack after millisecond flash lamp annealing at 400°C, 90 J/cm$^2$ and wet etch of the top electrode using $H_2O_2$. The thick $SiO_2$ layer limits capacitive contributions from Si. b) GIXRD spectrum at ω = 0.9°: the W bottom electrode is in the α-W phase. No monoclinic hafnia is found (Materlik *et al.* [12] and Muller *et al.* [13]).

Grazing Incidence X-Ray Diffraction confirms the crystallization of the $HfO_2/ZrO_2$ nanolayers in the orthorhombic or tetragonal phase [12], [13] with no fraction of monoclinic phase (**Fig. 2**). Third, a 1 µm thick thermal $SiO_2$ layer and an optimized design were used to prevent parasitic contributions in capacitance measurements. A 1 μm² FeCap cross-section (**Fig. 3**) shows the agreement between nominal and effective lateral dimensions.

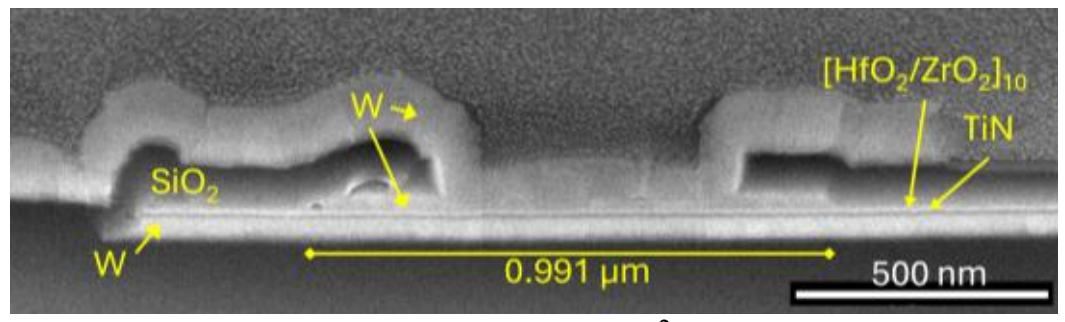

Fig 3. FIB-SEM cross-section of a 1 μm² device showing agreement between nominal and experimental dimensions.

## EXPERIMENTAL RESULTS

### *A. Multilevel Ferroelectric Switching*

Electrical characterization is performed on devices in the range of 0.12 to 1600 μm². A known drawback of device size reduction is the increase in device-to-device variability [14]: it rises to 20% for 1 μm² FeCaps at 1V (**Fig. 4a**). A remanent polarization of $P_r$ = 23 μC/cm² is achievable through full switching $P_{MAX}$ (**Fig. 4b**), suitable for memory storage.

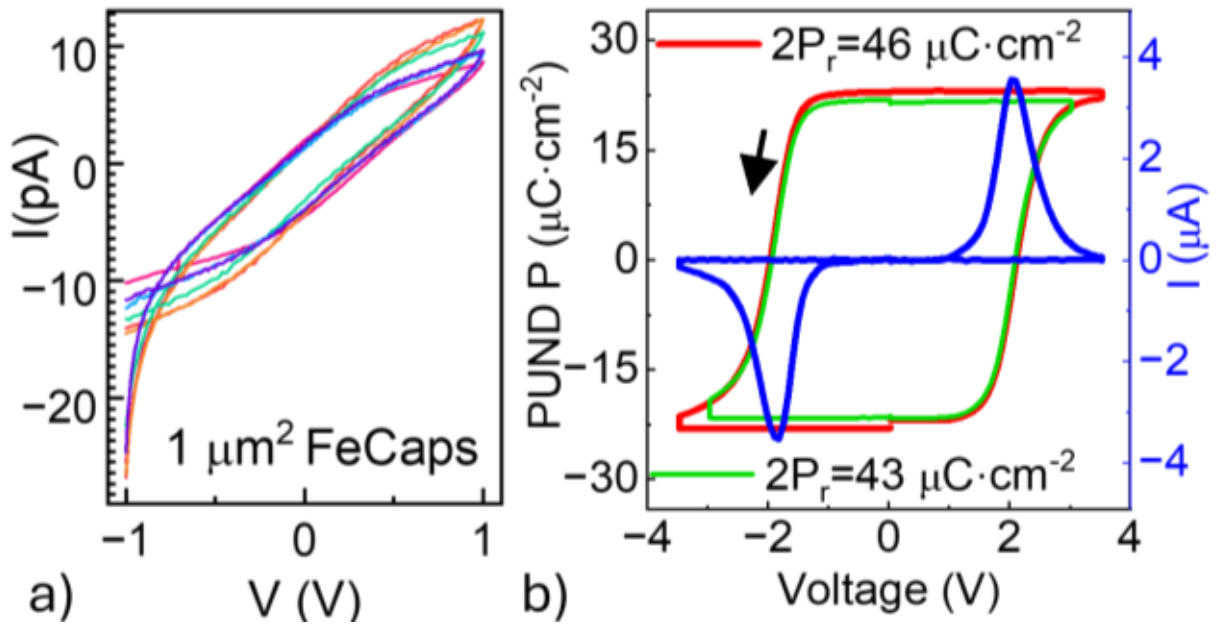

Fig. 4. a) Device-to-device variability in scaled devices. Positive-Up-Negative-Down (PUND) polarization and current on a 10x10 μm² FeCap showing robust switching and high remanent polarization.

To evaluate multilevel programming capability, a write-read scheme is used. We employ a modified Positive-Up-Negative-Down (PUND) Write-Read scheme consisting of a reset pulse, a programming pulse of amplitude $V_{write}$ and width $t_{write}$, and two read pulses (**Fig. 5a**).

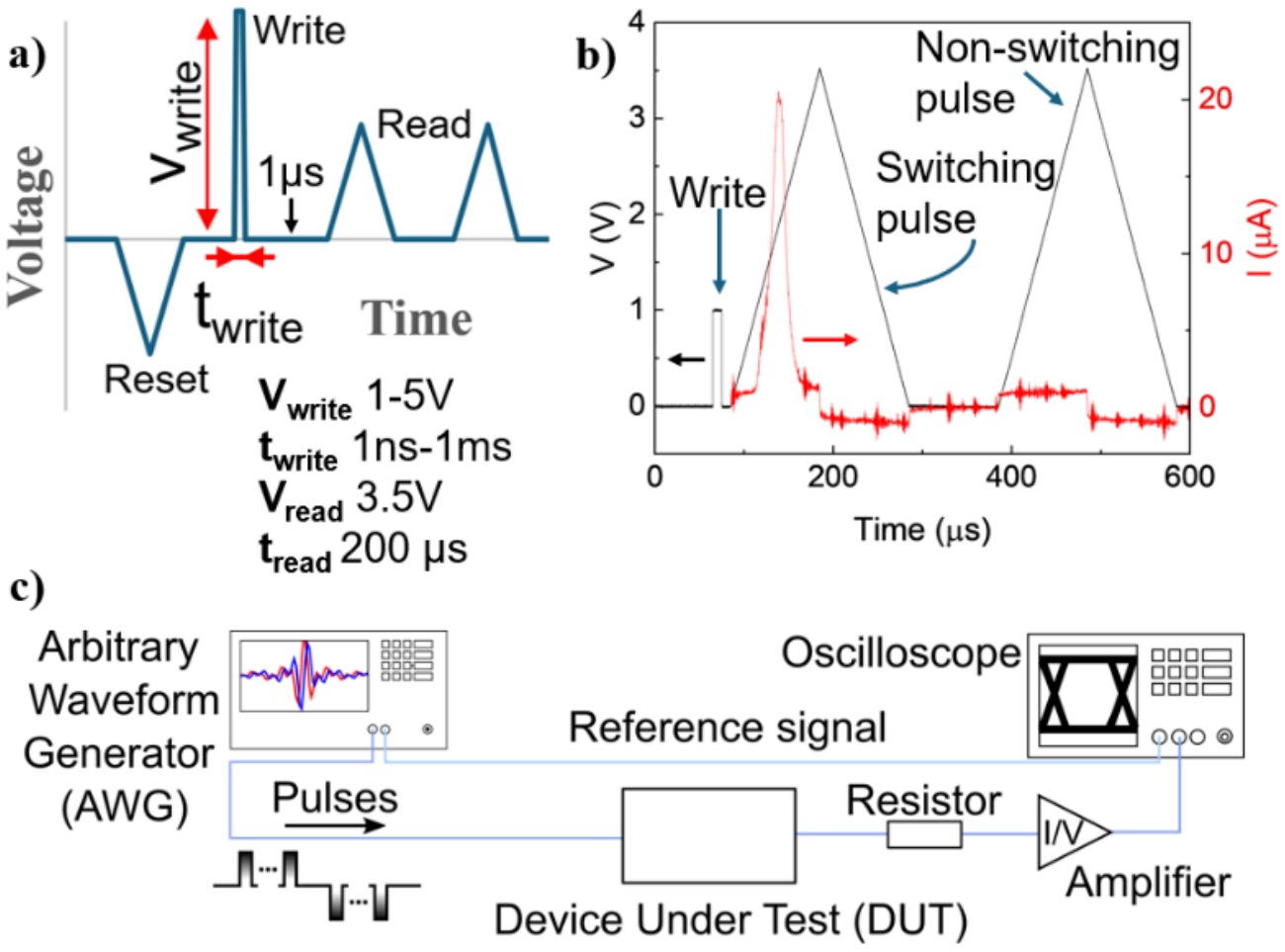

Fig. 5. a) Reset-Write-Read scheme for multilevel programming. b) V(t) and I(t) sweep showing the ferroelectric switching current (1st read pulse) and subsequent non-switching current (2nd read pulse). c) Setup for nanosecond programming and read.

The polarization P switching during the first read pulse "*P*" (destructive read) is quantified by the integral of the current in the first read pulse subtracted by the current in the second pulse "*U*" (**Fig. 5b**). The polarization charges and displacement currents scale with the FeCap area: to allow for the characterization of intermediate polarization states on submicrometer devices, the current is measured after amplification (**Fig. 5c**). The maximal polarization ($P_{MAX}$) is measured directly after a reset. The intermediate polarization states ($P/P_{MAX}$) in the 40×40 μm² FeCap on CMOS are accessible through the selection of different $V_{write}$ and/or $t_{write}$ (**Fig. 6**). The programming of the P states is in agreement with the Nucleation-Limited Switching (NLS) model [15], where $\boldsymbol{\tau_0}$ and $\boldsymbol{n}$ are fitting parameters :

$$\frac{\boldsymbol{P}}{\boldsymbol{2P_{MAX}}} = \int_{-\infty}^{+\infty} \left[\mathbf{1} - \mathbf{exp}\left\{\mathbf{1} - \left(\frac{\boldsymbol{t_{write}}}{\boldsymbol{\tau_0}}\right)^{\boldsymbol{n}}\right\}\right] \boldsymbol{F(log\tau_0)d(log\tau_0)}.$$

This validates that the quantified current arises from the change in FE polarization.

### *B. Nanosecond switching below 5V in scaled FeCaps*

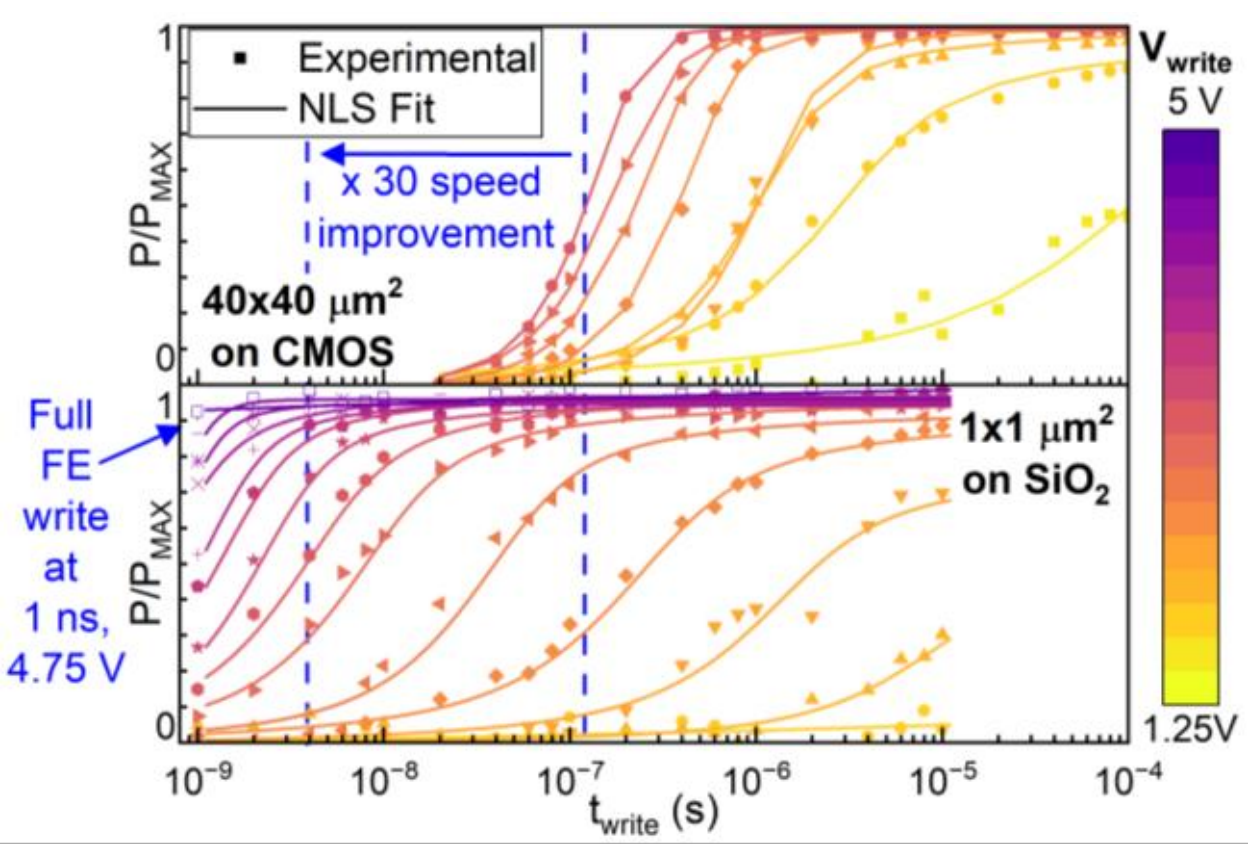

Fig. 6. Multilevel switching on CMOS accessible through control on $V_{write}$ or $t_{write}$. Device scaling allows full ferroelectric switching at only 1 nanosecond and $V_{write}$ < 5 V.

In polycrystalline hafnia, the FE switching time decreases with the device area [16], [17]. Scaled test structures down to 400 nm were fabricated on thermal $SiO_2$ to increase the programming speed. On a 2×2 μm² device, the $P/P_{MAX}$ intermediate levels at a given $V_{write}$ are programmed using 20 times shorter pulses compared to 40×40 μm² FeCaps.

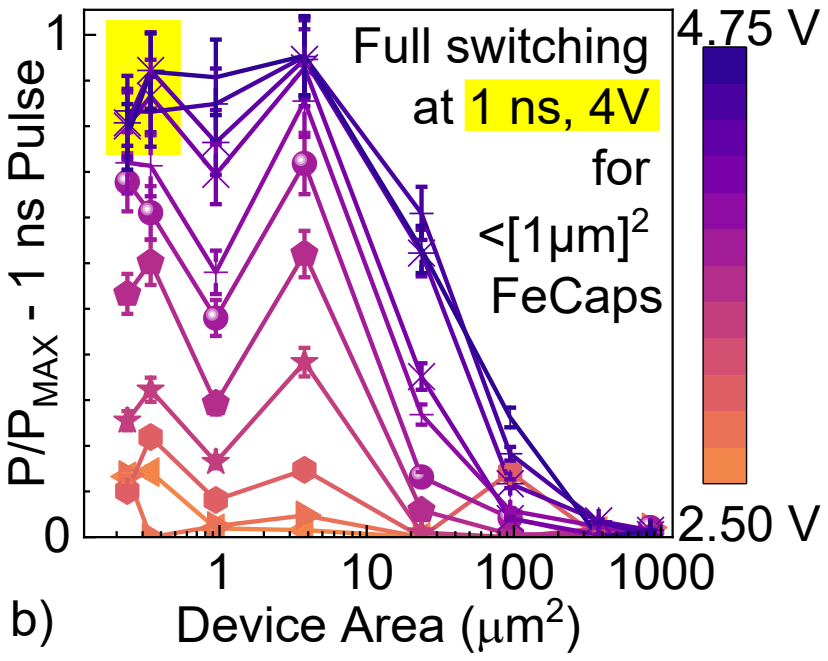

Fig. 7. Effect of scaling on nanosecond switching. The cycle-to-cycle variability falls below the instrumental error.

In the positive (slow) polarity, a record $V_{write}$<5 V is sufficient to fully switch a 2×2 μm² FeCap with a positive 1-

nanosecond pulse, which compares to state-of-the-art FeFETs (W=L=240 nm) from GlobalFoundries requiring +6V [17].

Further, in submicrometer devices, nanosecond switching requires less than 4V (**Fig. 7**). This improvement compared to FeFETs / FeCaps with a dielectric interlayer is explained by the metallic nature of the $WO_x$ oxide interlayer, in which the voltage drop is negligible.

### *C. MemCapacitive Device Operation*

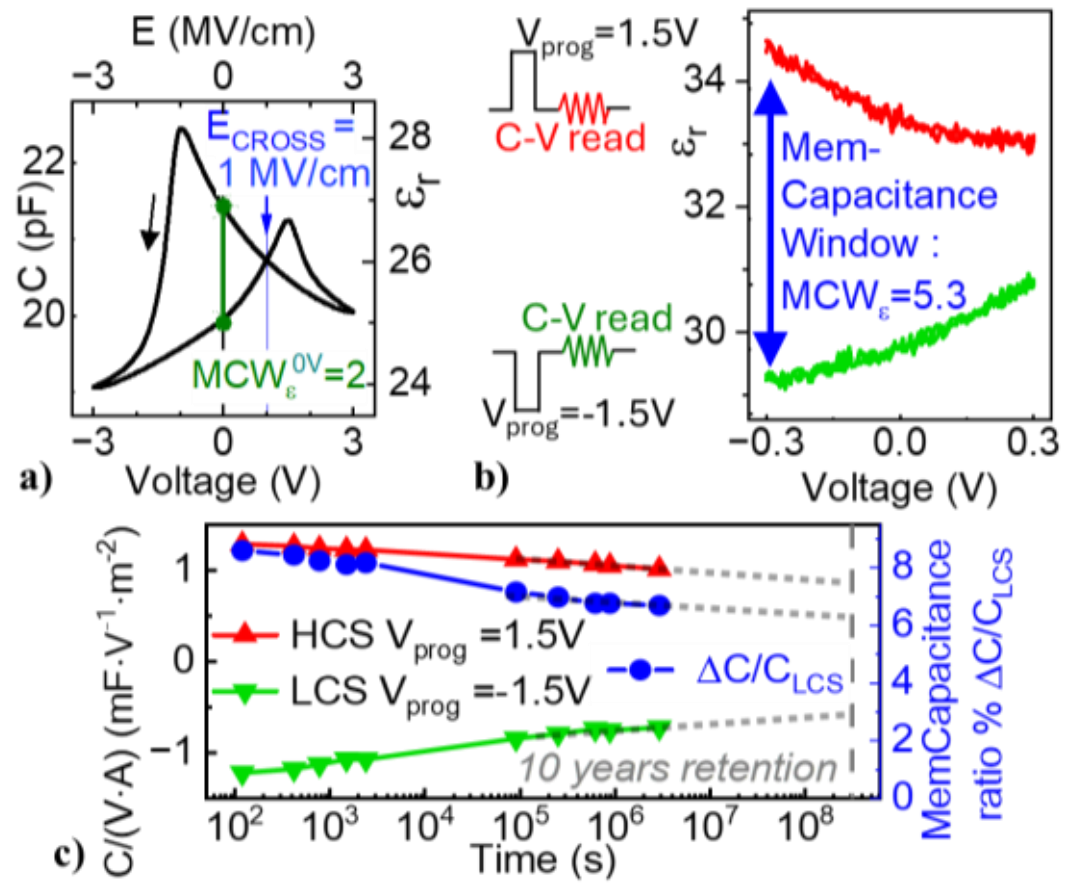

Fig. 8. a) The asymmetric FeCap exhibits a "butterfly" C-V loop with a crossover field $E_{CROSS}$=1 MV/cm estimated for a HZO thickness of 10 nm. b) The High and Low Capacitive States (H, LCS) are measured non-destructively by a C-V sweep at sub-switching voltages: partial FE switching at +/- 1.5 V DC opens a MC window. c) The MC ratio and C/V ratio project a 10-year retention.

The destructive nature of "switching current read schemes" limits, however, device endurance, even in scaled FeCaps operating at ultra-low voltages [18]. From this perspective, we investigate non-destructive read schemes, in particular MemCapacitance windows (MCWε) [5]. Thanks to the asymmetric materials stack, partial switching of the FE domains (± 1.5 V) is enough to modulate the relative permittivity $\varepsilon_r$ of the FeCap (**Fig. 8a**). The MC is measured non-destructively through C-V profiling at sub-switching voltages (**Fig. 8b**). The retention of MC is measured non-destructively at |V| < 0.3 volts for one month. An MC ratio > 6% is projected after 10 years retention (**Fig. 8c**), which is compatible with the requirements of neural network computation using memcapacitive crossbar arrays [5], [19].

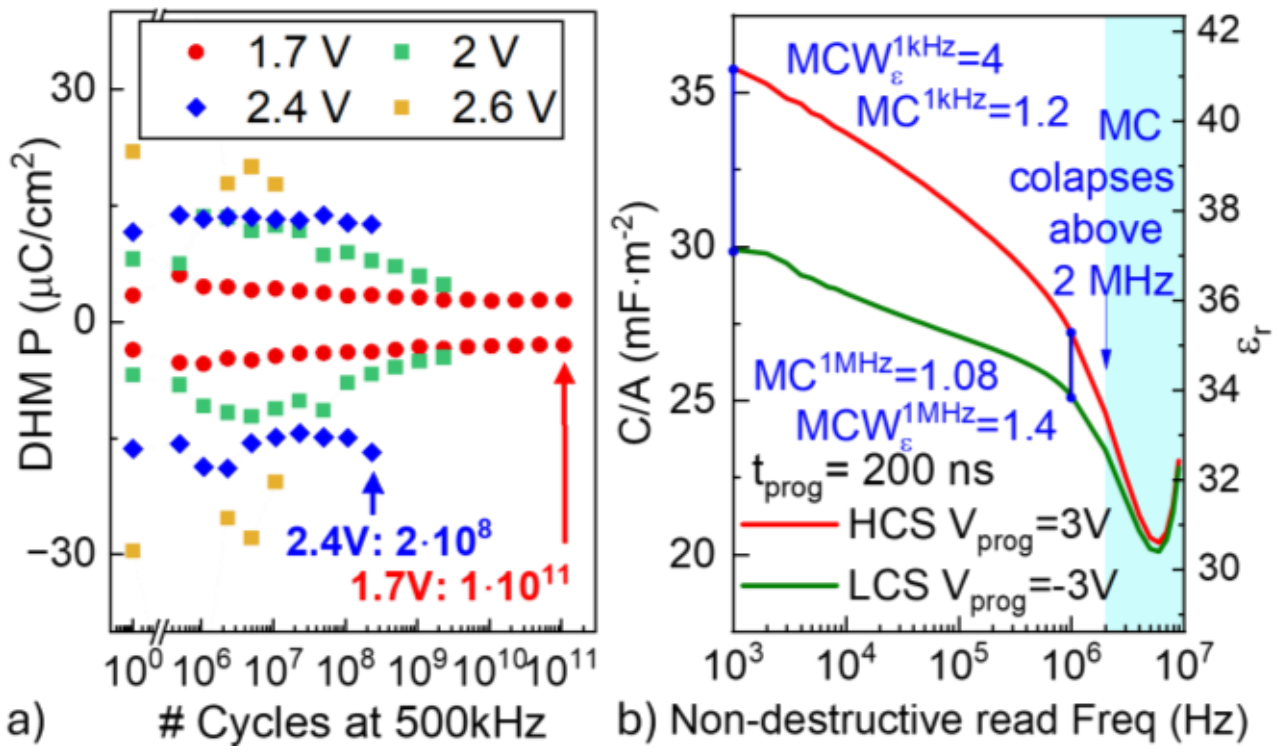

Fig. 9. a) Dynamic-Hysteresis-Mode (DHM) polarization fatigue. No wake up is needed. Partial switching at 1.7 V enables > 1.1E11 cycles. b) Non-destructive C-V read (+/- 300mV) at increasing read frequency, after programming in the HCS and LCS. The MCW collapses above 2MHz.

Programming the devices with partial polarization levels allows for an increase of their write endurance [12], achieving $2{\cdot}10^8$ cycles at ± 2.4 V and up to $1.1{\cdot}10^{11}$ cycles at ± 1.7 V (**Fig. 9a**). Such high endurance has only been reported for non-destructive read operations in MC devices [4, 14]; it is demonstrated here for MC programming operation. In addition, using the same scheme, we achieve non-destructive read speeds up to ten times faster than those from state-of-the-art MC devices [5, 15], up to 1 MHz (**Fig. 9b**).

### *D. 20-Picosecond Non-Destructive Read*

However, above read frequencies of 2 MHz, the MC collapses (**Fig. 9b**), limiting the operational speed of the potential memory technology. To address this limitation and enable high-speed applications, a charge-based non-destructive read scheme using electrical pulses was proposed [6]. Here, we explore the FeCaps' response at timescales shorter than the device's capacitive response, proposing a novel and non-destructive read procedure. We design an experiment at ultra-fast speeds to deconvolute the contributions from capacitance and switching, and prove that the capacitive and resistive responses are modulated by the polarization. The P state of a 2×2 µm² device is pre-set with a Source-Measurement-Unit where the switching current is fully observed. In the ultrafast setup (**Fig. 10a,** adapted from ref. [22]), a PUND read scheme is applied for varying voltages and pulse widths (**Fig. 10b**).

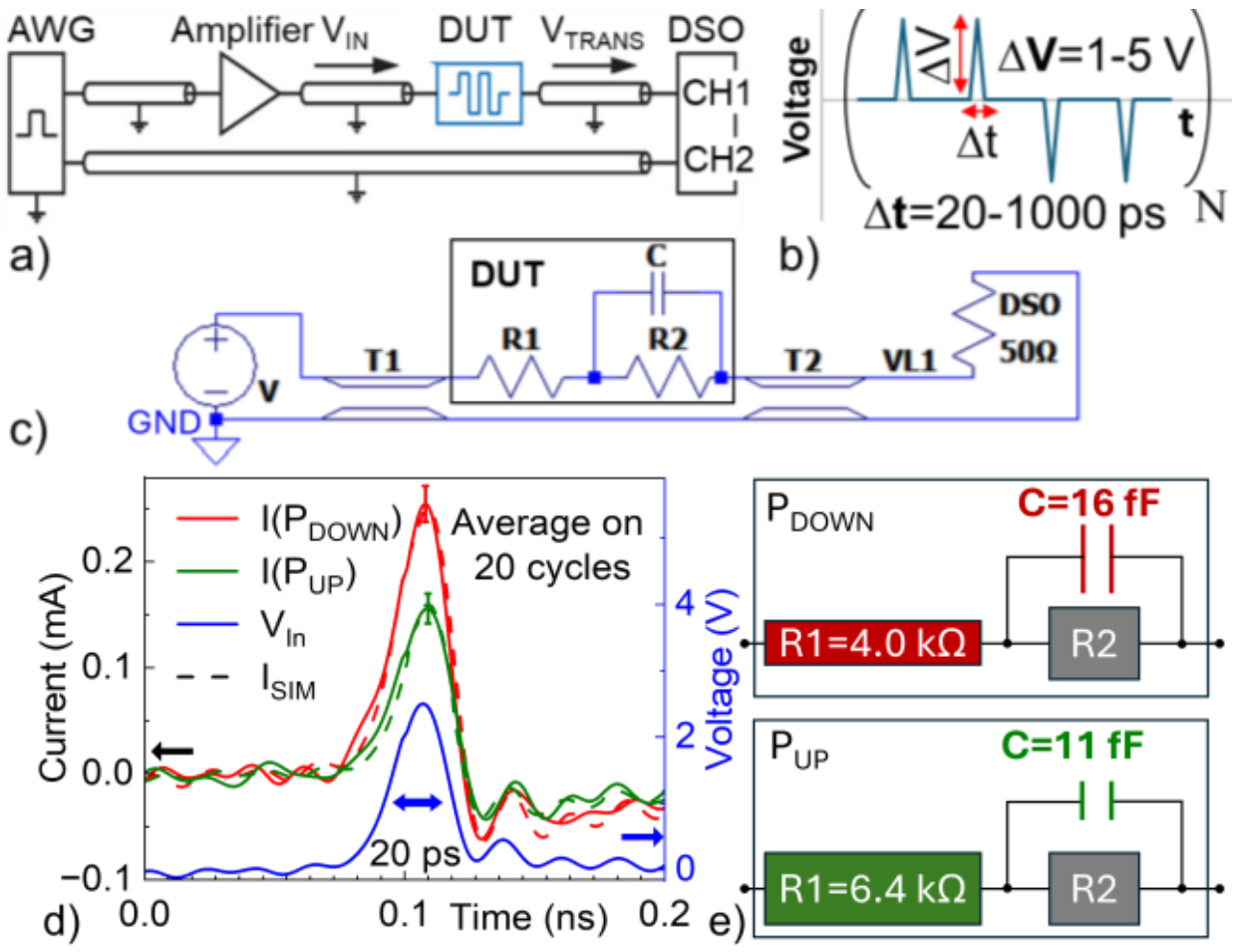

Fig. 10. a) Set-up for FeCap characterization under 20 ps pulses and b) PUND sequence. DUT: Device Under Test. T1,2: Transmission lines DSO: Digital Storage Oscilloscope
c) Leaky capacitor model implemented in SPICE.
d) Ultra-fast non-destructive read (experimental) of the two polarization states $P_{UP}$ or $P_{DOWN}$ using 20 ps pulses (left axis). SPICE simulation of the current response ($I_{SIM}$, left axis) to the experimentally measured voltage pulse (right axis), using the R1 and C values reported in e).

A leaky capacitor model [23] implemented in SPICE with three parameters (R1, R2, and C) (**Fig. 10c**) is used to simulate the current response for the two polarization states. For read pulses shorter than the RC time constant ($\tau_{RC}$) of the device (< 300 ps), the FeCap exhibits a purely resistive response. A High and a Low Resistive State are observed depending on the programmed polarization ($P_{UP}$ or $P_{DOWN}$) (**Fig. 10d**), similar to those previously reported on thin HZO nanolayers with a $WO_x$ functional layer, where the resistive response is dominant at every timescale [24]. This work reports, for the first time, an energy-efficient and non-destructive read of a FeCap operating in the purely resistive regime, dissipating approximately 14 femtojoules using 20-picosecond pulses. An optimal fit between simulated and experimentally measured $I(P_{UP})$ and $I(P_{DOWN})$ currents is achieved iteratively with the least-squares method to determine ($R1_{UP}$, $C_{UP}$) and ($R1_{DOWN}$, $C_{DOWN}$) for 20-ps pulses (dashed lines in **Fig. 10d** and **Fig. 10e**). The results are independent of R2 > 0.5 MΩ.

| | [25] **2022 VLSI** | [6] **2023 IEDM** | [26] **2024 IEDM** | [18] **2024 VLSI** | [11] **2024 EDTM** | [20] **2025 IMW** | **This work** |
|---|---|---|---|---|---|---|---|
| CMOS integration | n+ doping | BEOL compat. | BEOL compat. | BEOL compat. | BEOL | BEOL compat. | BEOL |
| $P_{MAX}$ Program. | 1 µs 6V | 50 µs 2.7V | 900 ns 3.3V | 1 µs 0.5V | 0.2 µs 3V | 1 µs 1.5 V | **1 ns 5 V** |
| Destructive Read | No | No | No | Yes | Yes | No | No |
| Mem-Capacitance ratio | 125 | 1.29 | 2.6 | No | - | - | 1.19 |
| Retention | 10 years | 10 years | - | 10 years | - | 10 years | 10 years |
| Read time | 10 µs | 500 ns | 500 ns | 1 µs | 200 µs | 1 ns | **20 ps** |

TABLE I. Benchmark

As shown in **Table I**, the proposed FeCap achieves faster switching at 5V thanks to materials and device design. The novel non-destructive read scheme leverages the modulation of the FeCap resistive response upon ferroelectric switching, allowing higher frequency read.

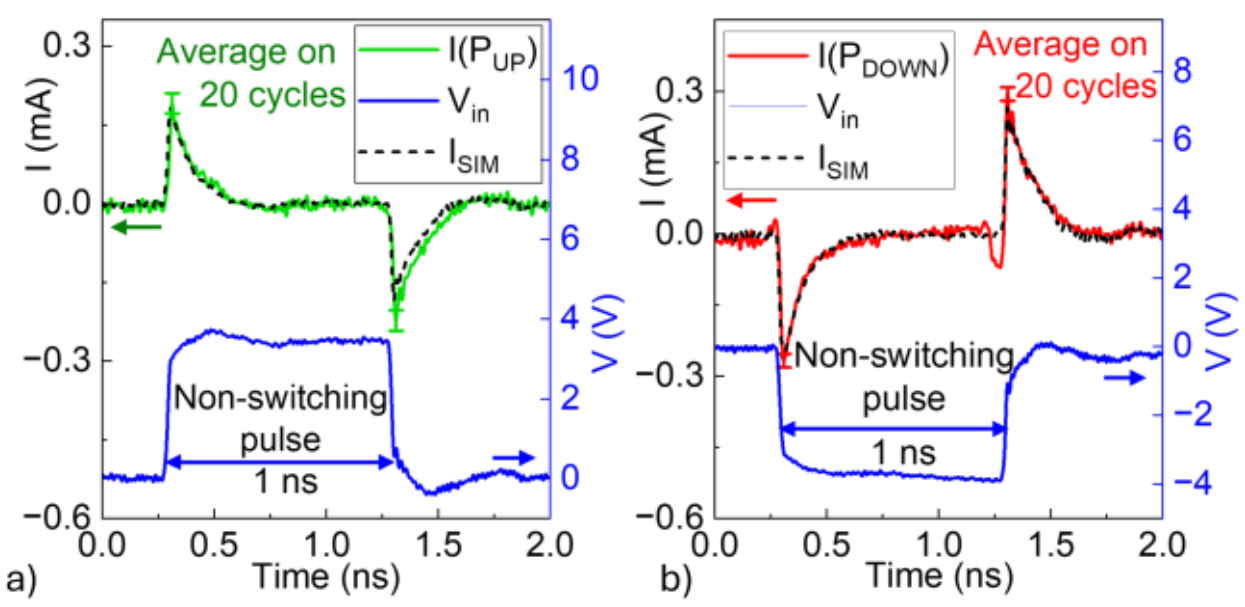

Fig. 11. A 1 ns non-switching pulse (blue lines) with rise time of 20 ps is applied to the FeCap. The resulting current exhibits the capacitive response for positive (a) and negative (b) bias. Cycle-to-cycle variability: the response is measured over 20 cycles (error bars in a) and b)). The response to the voltage pulse (right axis) is simulated with SPICE to fit the average current responses.

Similar to recent reports by Dahan *et al.* [20], the non-destructive read can also be achieved above $\tau_{RC}$, for which the capacitive response is observed. The experimental data at 1 ns (**Fig. 11**) are used to refine our leaky capacitor models in the UP and DOWN states. Using the experimental uncertainty as a tolerance band in the TLS fitting, we find a set of parameters ($R1_{UP}$, $C_{UP}$) and ($R1_{DOWN}$, $C_{DOWN}$) that fit for both the capacitive (1 ns) and resistive (20 ps) responses (**Fig. 12**). This confirms that despite the comparatively small On/Off ratio (~1.5), a single unipolar 20 ps pulse is sufficient to infer the P state of the device.

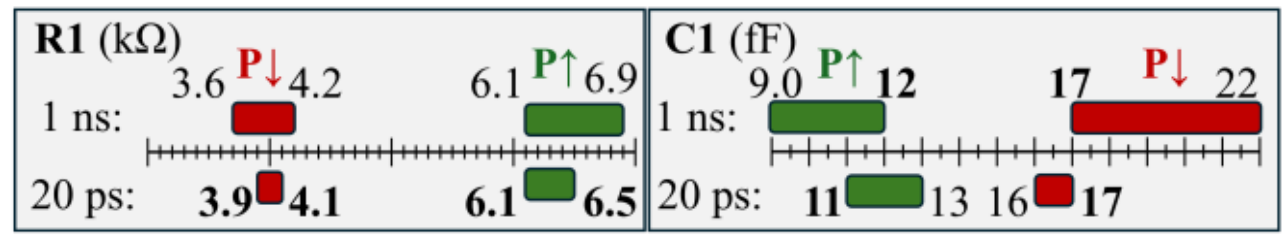

Fig. 12. Confronting the SPICE simulations to experimental data at 20 ps and 1 ns allows to narrow the model parameters and reveal the origin of the memory window: as the polarization is switched upwards, C decreases from C=17 fF to C ϵ [11-12 fF]. R1 increases from R1 ϵ [3.9-4.1 kΩ] to R1 ϵ [6.1-6.5 kΩ]. The two states are distinguishable with a 20 ps non-destructive read.

## SUMMARY

We demonstrated ferroelectric $HfO_2/ZrO_2$ nanolayers integrated in the BEOL of CMOS displaying multilevel programming below 5 volts. Nanosecond switching was further achieved in micrometer-scale devices on Si at 4 V. Partial FE switching allows over $10^{11}$ write cycles endurance with a projected memcapacitance ratio of 6% at 10 years. Overcoming the collapse of the MC for read frequency above 2 MHz, we proposed an ultrafast (20 ps), non-destructive, energy-efficient (14 fJ) read method based on sub-$\tau_{RC}$ pulses and explained the FeCap response using an equivalent circuit.

**ACKNOWLEDGMENTS**: We thank the BRNC, S. R. Mamidala, D. F. Falcone, U. Drechsler, A. Olziersky, & S. Reidt at IBM. **Funding**: SNSF ROSUBIO #218438 & ALMOND #198612, SERI SwissChips, EU ViTFOX #101194368, CogniGron, Ubbo Emmius Funds (Uni. of Groningen)